\documentclass[useAMS,usenatbib,usegraphicx]{mn2e}
\def\etal{\textit{et al.}\ }

\usepackage{soul,color}
\usepackage{float}
\usepackage{placeins} 
\definecolor{lightgray}{gray}{0.77}
\sethlcolor{lightgray}

\title[Cluster Membership]{Cluster Membership Probability: Polarimetric Approach}
\author[Biman J. Medhi]
{Biman J. Medhi$^{1}$\thanks{E-mail: biman@aries.res.in}, Motohide Tamura$^{2}$\\
$^{1}$Aryabhatta Research Institute of Observational Sciences, Manora Peak, Nainital - 263 129, India \\
$^{2}$National Astronomical Observatory of Japan, Mitaka, Tokyo 181-8588, Japan \\}
 
\begin{document}

\date{}

\pubyear{2012}

\maketitle

\label{firstpage}

\begin{abstract}

Interstellar  polarimetric data  of the six  open clusters
Hogg 15, NGC 6611, NGC 5606, NGC  6231, NGC 5749 and NGC 6250 have been
used  to estimate the membership  probability for the stars within them.
For proper-motion member stars, the membership  
probability estimated using the polarimetric data is
in good agreement with the proper-motion cluster membership probability. 
However, for proper-motion non-member stars, the membership probability estimated by the polarimetric 
method is in total disagreement
with the proper-motion cluster membership probability.
The inconsistencies in the determined memberships may  be because  of  the  fundamental
differences between  the two  methods of determination: one is  based on
stellar proper-motion in space and the other is  based on selective  extinction of
the stellar  output by the  asymmetric aligned dust grains  present in
the interstellar medium. The results and analysis suggest that
the scatter of the Stokes vectors $q(\%)$ and $u(\%)$
for the proper-motion member  stars depends on the interstellar
and intra-cluster   differential reddening in the open cluster. 
It is found that this method could be used to estimate the cluster membership probability 
if we have  additional  polarimetric and photometric
information  for a star to identify it as a probable member/non-member of a particular cluster, such as    
the   maximum wavelength   value  
($\lambda_{max}$),  the unit  weight  error of  the fit  ($\sigma_1$),
the dispersion  in the polarimetric  position  angles ($\overline{\epsilon}$), 
reddening ($E(B-V)$) or the  differential intra-cluster reddening 
($\Delta E(B-V)$). This method could also be used to estimate
the membership probability  of known member stars having no  membership probability 
as well as to resolve disagreements about membership among different 
proper-motion surveys (Dias $\etal$2006, Baumgardt $\etal$2000, 
Belikov $\etal$1999, Tucholke $\etal$1986, Berger 1982).

\end{abstract}


\begin{table*}
\centering
\begin{minipage}{170mm}
\caption{List of selected open clusters for analysis}\label{list_oc}
\begin{tabular}{llllll}
\hline
Cluster Name & {\hspace {-2mm}$E(B-V)$(in mag)} & Distance (in kpc) & Age (in Myr.) & \ \ \ \ Polarimetric data & Proper-motion membership \\
{\hspace {5mm}(1)} & {\hspace {8mm}(2)} & {\hspace {6mm}(3)} & {\hspace {5mm}(4)} & {\hspace {12mm}(5)} & {\hspace {12mm}(6)}\\
\hline
Hogg 15  & \ \ \ 1.16 $\pm$ 0.03 & \ \ \ 3.0 $\pm$ 0.3 & \ \ \ 6.0 $\pm$ 2 & \ \ \ \ Orsatti \etal 1998 & \ \ \ \ \ Dias \etal 2006 \\
NGC 6611 & \ \ \ 0.85 $\pm$ 0.05 & \ \ \ 3.2 $\pm$ 0.3 & \ \ \ 3.0 $\pm$ 2 & \ \ \ \ Orsatti \etal 2000  & \ \ \ \ \ Belikov \etal 1999 \\
NGC 5606 & \ \ \ 0.50 $\pm$ 0.05 & \ \ \ 2.4           & \ \ \ 6.0 $\pm$ 2 & \ \ \ \ Orsatti \etal 2007   & \ \ \ \ \ Dias \etal 2006 \\ 
NGC 6231 & \ \ \ 0.46 $\pm$ 0.05 & \ \ \ 1.6 $\pm$ 0.05& \ \ \ 4.0 $\pm$ 1 & \ \ \ \ Feinstein \etal 2003 & \ \ \ \ \ Dias \etal 2006 \\
NGC 5749 & \ \ \ 0.42 $\pm$ 0.04 & \ \ \ 1.2 $\pm$ 0.18& \ \ \ 27.0        & \ \ \ \ Vergne \etal 2007 & \ \ \ \ \  Dias \etal 2006 \\
NGC 6250 & \ \ \ 0.33 $\pm$ 0.05 & \ \ \ 1.0           & \ \ \ 14.0        & \ \ \ \ Feinstein \etal 2008 & \ \ \ \ \ Dias \etal 2006 \\ 
\hline
\end{tabular}
\end{minipage}
\end{table*}

\section{Introduction}

In studies of star clusters there are 
several interesting aspects to be understood, such as stellar evolution,  galactic structure and
evolution,  and stellar  dynamics.   Compared to single  stars,
distant star  clusters can  be identified more  easily and their  age can
be determined more reliably. Since their inception, studies of star 
clusters have focused on cluster membership.

Proper-motion   studies  on   cluster  membership  have made  very
significant contributions to star cluster  research. The
basic  goal of   astrometric cluster  membership  studies is  the
production  of a  color-magnitude  diagram of  probable members  with
reduced field star contamination (Cudworth 1997). It is difficult to confirm or discount
the    membership   of   stars    having     peculiarities, $e.g.$,
pre-main-sequence  stars,  super-giants,  Cepheids or  other  variables
(Cudworth 1997). Proper motions with  standard errors of 4 to 7 mas
$yr^{-1}$  will convey  membership  information at  magnitudes down  to
$\sim$ 16th magnitude (Zacharias $\etal$2004). Moreover, such precision
can be achieved with more than  one image/plate at each epoch and an
epoch difference  of 10 to 15  years depending on the  distance to the
object when the  telescope plate scale is $\sim$  10 arcsec $mm^{-1}$ (Cudworth 1986,1997).
However,  several disagreements about memberships among different
proper-motion surveys have arisen (Dias $\etal$2006, Baumgardt $\etal$2000, 
Belikov $\etal$1999, Tucholke $\etal$1986, Berger 1982). To overcome the 
limitations of the current method, another robust method is 
required that is independent of the current method.

The light from distant stars is partially plane 
polarized, which is thought to
be due to dust  grains in  the interstellar medium,
which are also responsible for  the reddening of  starlight. According to  the Davis and
Greenstein mechanism, the polarization of
starlight is caused by   selective extinction due to asymmetric
dust grains  aligned in the interstellar  medium, possibly by the galactic
magnetic field  (Davis $\etal$1951).   However, identifying the  dominant grain alignment
mechanism  has proved  to  be  an  intriguing problem  in  grain  dynamics
(Lazarian $\etal$1997). If the polarization is specifically produced by
the dust  grains present  in the interstellar  medium, then it will  
depend on distance as well as the generation method  of the dust grains
in  that  line  of  sight.  Hence,   the  percentage  of
polarization and position angle  along with the interstellar reddening may
provide an  independent measure  of cluster membership probability  under certain
conditions (Feinstein $\etal$2008, Vergne $\etal$2007, Berger 1982).  
In this paper, the consistency tests,  assumptions and
validity of polarimetric cluster membership probability in comparison with
proper-motion cluster membership probability will be explored using interstellar
polarimetric data available for different open clusters.
The rest of the paper is organized as follows. In Section 2 we present the concept 
of polarimetric cluster  membership and
selection of open clusters. The details of the procedures and  method 
adopted for estimating polarimetric cluster membership probabilities are
presented in Section 3. In Section 4 we present the scatter of the Stokes vectors 
in q(\%) versus u(\%) plots. In Section 5 we discuss intrinsic 
sources of polarization in stars. The polarimetric cluster membership probability for stars 
is presented in Section 6. In Section 7 we present a detailed study of the open cluster
NGC 6231. Finally, a discussion is presented in Section 8, and we conclude with a summary in Section 9.

\section{Polarimetric cluster membership and selection of open clusters}

In the past few years,  interstellar polarization has been  used to
obtain the cluster membership  probability for individual stars in different open clusters
(Feinstein $\etal$2008, Vergne $\etal$2007, Berger 1982). 
The  stellar output of individual  stars of different
clusters passing  through a  substantial amount of  interstellar matter
is subject to extinction and linear polarization. Both phenomena depend on
the particle size distribution of the aligned dust grains and vary as
a  function of  the  product  of the particle  size  distribution and  the
appropriate  cross section for  extinction and  polarization.
However, the polarization also depends on the fraction
of asymmetric dust grains of a particular size  which are aligned by the galactic magnetic field.
A correlation between these two phenomena cannot be obtained  for all
cases due  to  variations of the  grain alignment  efficiency.
Entire populations of  unaligned and unelongated grains may
contribute to extinction but not  to polarization. 
Ideally, the
member  stars of a particular cluster  should show  similar interstellar
polarization and  position angles because their light outputs  encounter the same amount 
of dust grains and a homogeneous magnetic  field, as they are located  at nearly the same distance. 
It is expected that non-member stars will show different interstellar polarizations
and position angles because they are located at different distances/lines of sight and their light outputs encounter 
different amounts and sizes of dust grains.
However, this  may not be true  for all cases.
If the distribution  of  dust grains and  
magnetic fields are not uniform inside a cluster and/or in a line of sight,  
then the  member stars of a cluster would show
different  interstellar  polarizations  and position  angles. Large scale changes
in the dust grain distribution and magnetic field homogeneity in different parts 
of the line of sight may also cause depolarization of initially polarized light.
Moreover,
it is possible that  some of the member stars may have an
intrinsic source  of linear polarization.  In that  case, the intrinsic
component of polarization  may  enhance or depolarize  the  interstellar component of polarization
of  that particular star.   To apply the method successfully to
estimate the cluster membership probability based on interstellar
polarization,  the  amount   of  interstellar selective
extinction  and polarization  vector  should be  similar  for all  the
member stars of a cluster.

To test the consistency of this 
technique we selected six open clusters for analysis
based  on the  following criteria:  (1)  available interstellar
polarization   data,   (2)   existence  of  a proper-motion  cluster  membership
probability  and (3)  all  samples distributed  over a  wide
coverage  of  reddening.  The  third  criterion  is set to  check  the
dependency of  polarization upon reddening. 
  The open  clusters Hogg 15,
NGC 6611, NGC 5606, NGC 6231,  NGC 5749 and NGC 5606 fulfill the above
criteria  with a the  reddening coverage from  1.15 to  0.37. A
brief description of the clusters follows and the important 
parameters of the clusters are given in Table $\ref{list_oc}$.

The $6 \pm 2$-Myr-old highly reddened  open cluster Hogg 15 is located  at a distance of
$3 \pm 0.3$ kpc (Sagar $\etal$2001).  It
is one of the few clusters known to lie in the second inner arm of our
Galaxy. Hogg 15 is effected  by non-uniform reddening across the cluster. 
The differential and average values of
the reddening are nearly 0.20 mag and  $1.16 \pm 0.03$  mag, respectively (Sagar $\etal$2001, Moffat 1974). 
Orsatti  $\etal$1998 have performed  a  multi-band
polarimetric study on 23 stars in Hogg 15. Of these 23 stars, only 17 have an available
proper-motion cluster membership probability (Dias $\etal$2006).
So, for analysis we  have taken the polarimetric data from Orsatti
$\etal$1998  and the proper-motion  cluster membership probability from  Dias $\etal$2006.

NGC 6611 is a very young open  cluster located at  a distance  of  
$3.2 \pm  0.30$ kpc and embedded in an ionized  hydrogen complex (M16)
in the Sagittarius spiral arm  (Winter $\etal$1997, Sagar $\etal$1979). 
The  extinction law  in the cluster  NGC 6611 is  variable. The
value of extinction found by different observers varies from $R_V =
2.5 \pm 0.6$ to $R_V =  3.4 \pm 0.7$ (Turner 1994, Sagar $\etal$1979, 
Gebel 1968, Johnson 1968). The average and 
differential values of reddening are nearly 0.85 $\pm$ 0.05 mag  and   0.63  mag,
respectively (Piatti $\etal$2002, Sagar $\etal$1979). 
IR studies concluded that the
variable extinction  in the north-west  area of the cluster  is caused
either by circumstellar or intra-cluster dust (de Winter $\etal$1997, 
Hillenbrand $\etal$1993, Chini $\etal$1983, Sagar $\etal$1979).
Orsatti $\etal$2000 have performed a multi-band polarimetric study on 39 stars
in this cluster. So, for analysis we have taken the polarimetric data  from  
Orsatti $\etal$2000 and  the proper-motion cluster membership probability 
for all 39 stars from  Belikov $\etal$1999.

The 6 $\pm$ 2-Myr-old open cluster NGC 5606  is located at a distance of $2.4$ kpc
(Piatti $\etal$2002, Vazquez  $\etal$1991). The reddening across the
cluster is variable. Differential and  average values of reddening are nearly 0.32 mag 
and 0.50 $\pm$ 0.05 mag, respectively (Piatti $\etal$2002, vazquez $\etal$1994).  
Orsatti $\etal$2007 have
made  multi-band polarimetric  observations  on  54  stars in  the
direction  of  NGC 5606. Of these 54 stars only 20 have an available proper-motion
cluster membership probability.   So,  for analysis, we  have taken  the
polarimetric   data  from   Orsatti  $\etal$2007  and
the proper-motion cluster membership probability from Dias $\etal$2006.

NGC 6231 is a  young open cluster located in the  core of the Sco
OB1 association at  a distance of $1.6 \pm 0.05$ kpc. The age of the cluster is nearly 4 $\pm$ 1 Myr
(Sana $\etal$2007, Piatti $\etal$2002). 
The average reddening of the cluster is $0.46 \pm 0.05$ mag and it is 
variable inside the cluster (Sung $\etal$1998). 
The value of differential reddening across the cluster is
nearly $0.28$ mag (Feinstein $\etal$1968). Feinstein $\etal$2003 have performed  a 
multi-band polarimetric study on 35 stars in the cluster. In  this paper we  use the
polarimetric  data from  Feinstein $\etal$2003  and the  proper-motion
cluster membership probability for all 35 stars form Dias $\etal$2006.

The 27-Myr-old, poorly  populated  open  cluster  NGC  5749  lies  near  the
south-western edge of the Lupus constellation  and is located at a distance of
$1.28 \pm 1.18$  kpc (Claria $\etal$1992). The average reddening of
the cluster is nearly $0.42 \pm  0.04$ mag(Claria $\etal$1992). The reddening across the
cluster is variable and the value of
differential reddening is nearly $0.13$ mag (Claria $\etal$ 1992).
Vergne $\etal$2007  have performed
multi-band  polarimetric observations  on 31 comparatively bright stars in the cluster NGC 5749.
Of these 31 stars, only 15 have an available proper-motion cluster membership
probability. So, for analysis,  we have taken  the  polarimetric data from Vergne $\etal$2007
and the proper-motion  cluster membership  probability from  Dias $\etal$2006.

The open cluster NGC 6250   lies  at the boundary of  the next
inner spiral (Sag-Car) feature and is located at a distance of $1.0$ kpc (Bayer $\etal$2000).  
It is effected  by
differential reddening across the cluster and the  values of differential and average 
reddening are nearly 0.28 mag and 0.33 $\pm$ 0.05
mag,  respectively (Bayer $\etal$2000, Herbst 1977). The  estimated age of the cluster is nearly $14$ Myr
(Bayer $\etal$2000). 
Feinstein $\etal$2008 have performed a multi-band polarimetric study on 32 stars in this cluster. 
Of these 32 stars 29 have a proper-motion cluster membership probability. 
So, for  analysis the polarimetric data and the proper-motion  cluster membership
probabilities are  taken from  Feinstein $\etal$2008 and Dias $\etal$2006, respectively.

\begin{figure*}
\begin{center}
\includegraphics[scale = .35, trim = 0 0 0 0, clip]{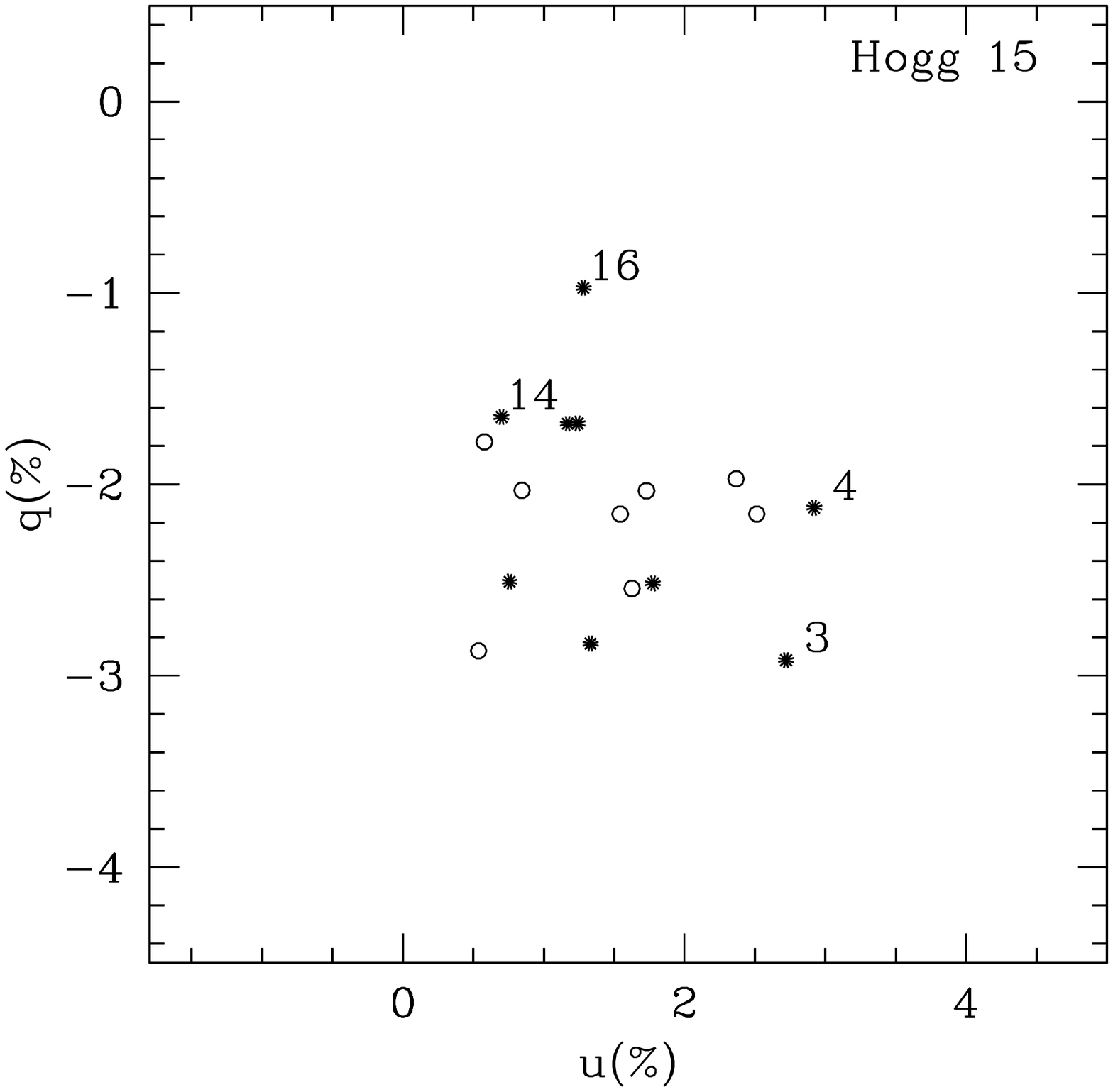}%
\hspace*{10pt}
\includegraphics[scale = .35, trim = 0 0 0 0,clip]{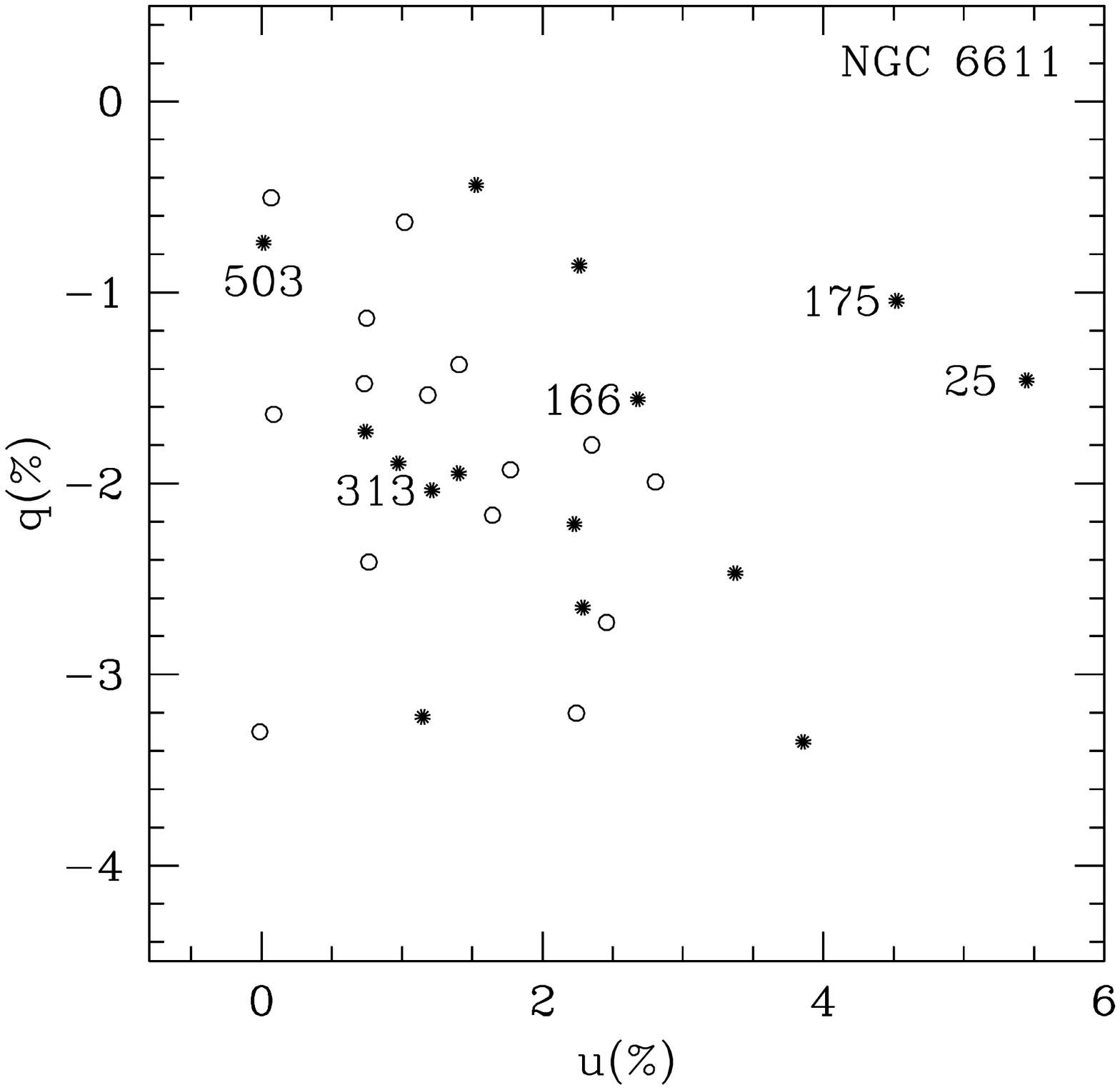}\\
\includegraphics[scale = .35, trim = 0 0 0 0,clip]{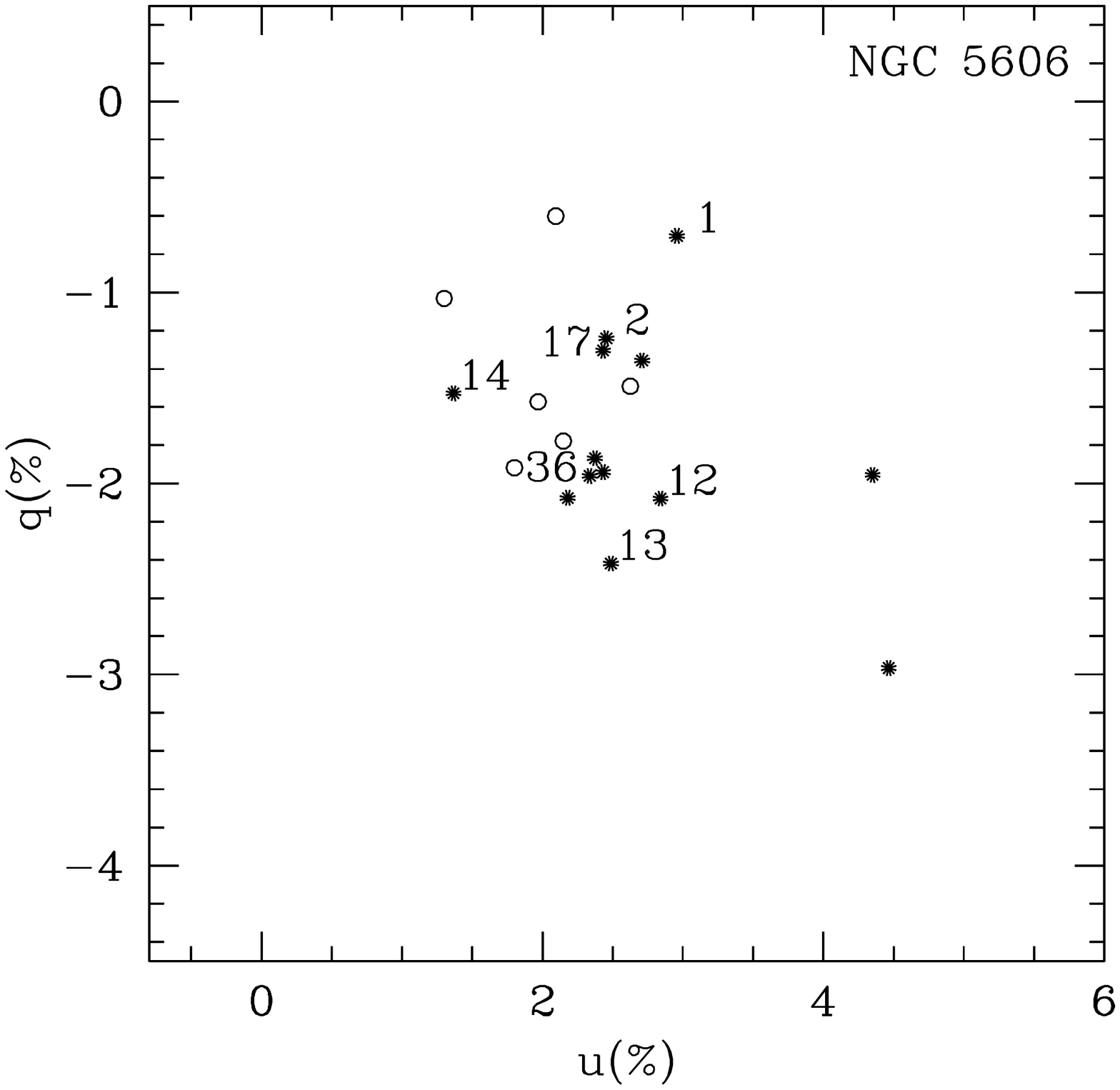}%
\hspace*{10pt}
\includegraphics[scale = .35, trim = 0 0 0 0,clip]{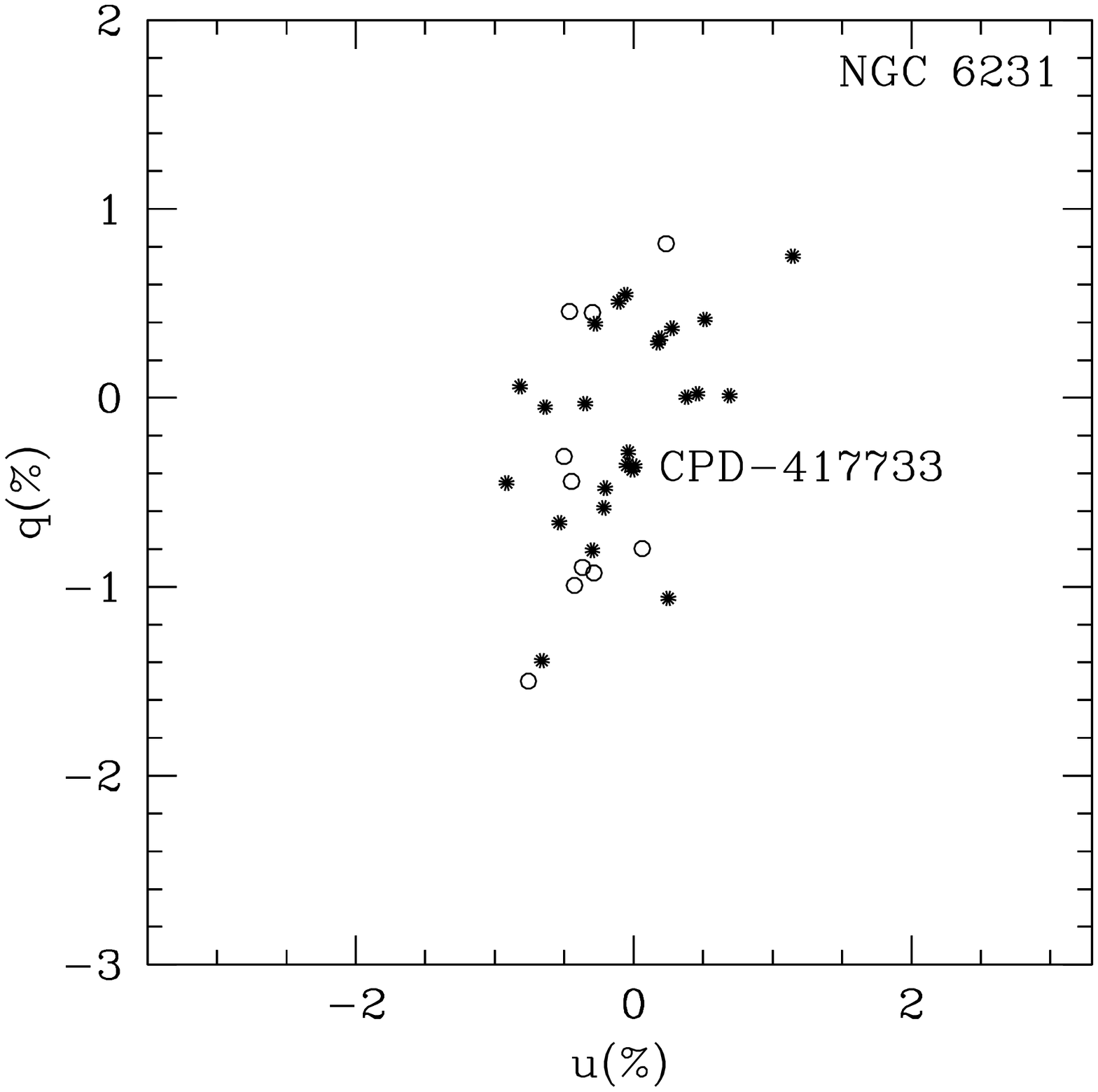}\\
\includegraphics[scale = .35, trim = 0 0 0 0, clip]{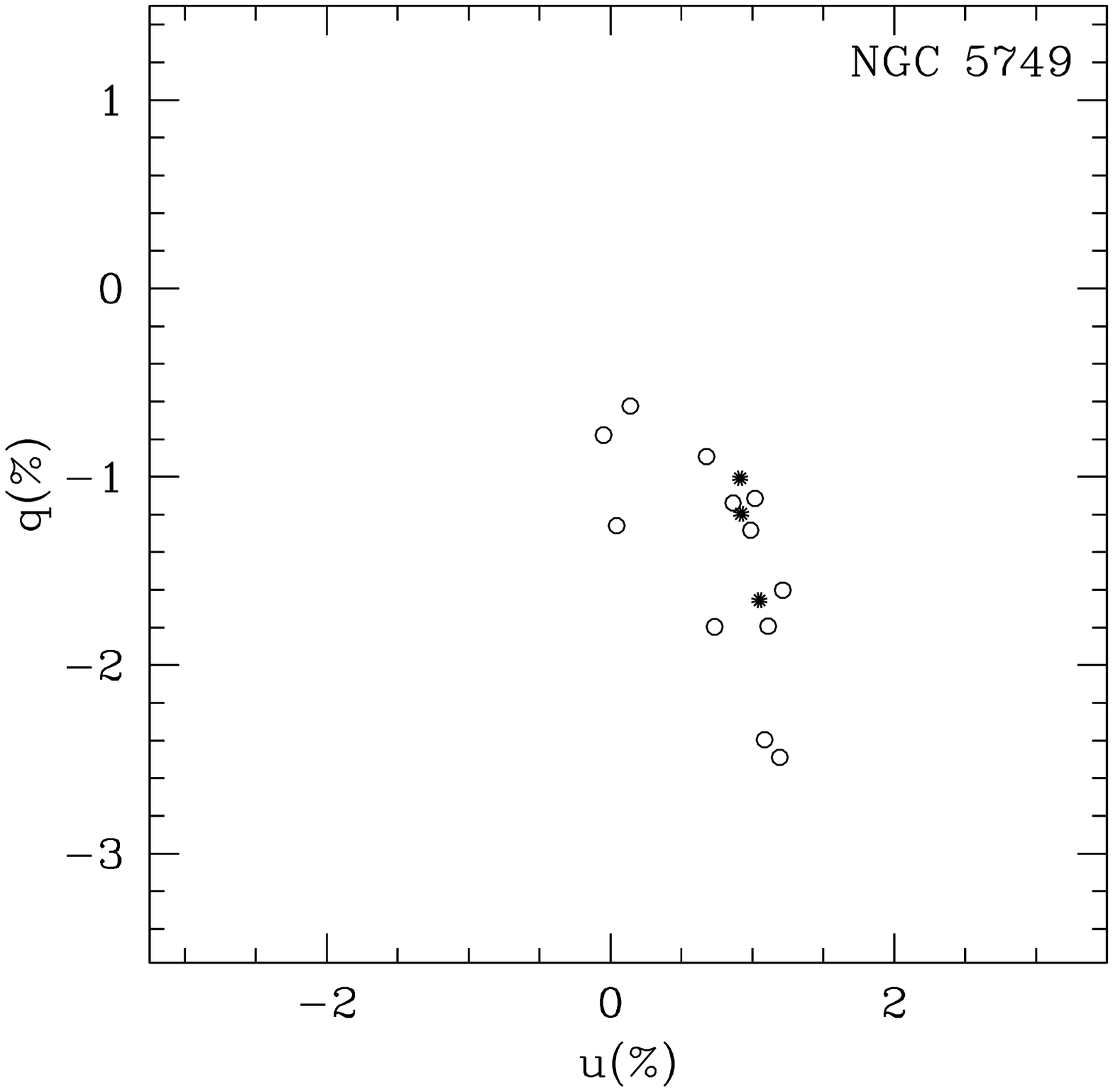}%
\hspace*{10pt}
\includegraphics[scale = .35, trim = 0 0 0 0, clip]{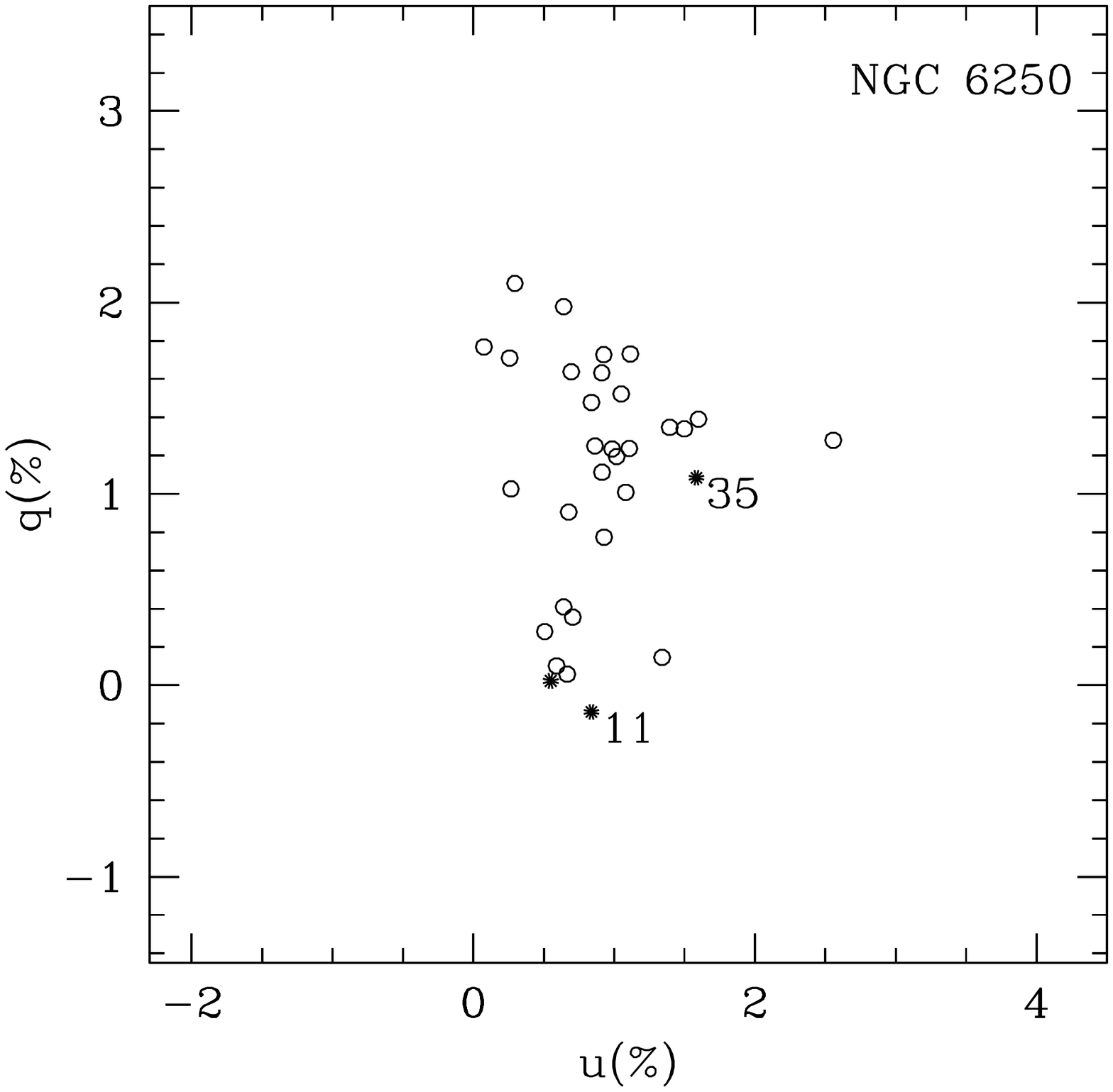}\\
\hspace*{10pt}
\caption[Q vs. U plots of different clusters]{ Polarization Stokes 
vectors $u(\%)$ and $q(\%)$ for stars with an available proper-motion cluster membership probability. 
Proper-motion cluster 
members are shown by filled-circle and non-members by open-circle symbols. The same range of X $(u(\%))$ and Y $(q(\%))$ scales 
are used in all the plots to visualize and compare the scatter in the different clusters.}
\label{fig_uq} 
\end{center}
\end{figure*}

\section{Estimation of the polarimetric cluster membership probability}

The proper-motion cluster membership probability can be used as a reference
for a consistency  test of the polarimetric cluster membership probability.
The stars of the different clusters were divided into four groups based
on the proper-motion cluster membership probabilities, $viz.$,
(1) proper-motion members  with  very high cluster
membership   probability, i.e. ,  $\ge 80\%$,  (2) proper-motion members  with a cluster
membership  probability between $50\%$ and $80\%$, (3) proper-motion non-members  with a cluster membership probability $<
50\%$ and (4) proper-motion non-members with very low cluster membership probability, i.e. , $<
20\%$ (Baumgardt $\etal$2000, Berger 1982).

The cluster membership  probability (${Mp_{pol}}^p$) is
estimated from  the average  deviation of the Stokes  vectors $q$  and $u$ of an individual star
from the mean values of $q$  and $u$ of the proper-motion group-one stars (groups of stars with very high proper-motion
membership probability $> 80\%$). 
We may consider the proper-motion
group-one  stars   as   representative  of  a particular  cluster  because  they have
a very high membership probability.
Percentage scaling/calibration of ${Mp_{pol}}^p$ is performed using
the full ranges (difference between maximum and minimum) of the Stokes vectors $q$ and $u$, 
considering the same scale of $100\%$ cluster membership probability. 
So, to estimate any individual star's 
cluster membership probability we compare the average deviation of the Stokes  vectors 
with these ranges.
The ranges of $q$ and $u$
are determined as being between the proper-motion group-one and group-four stars.

However, it is not possible to apply the same technique in a cluster where the proper-motion
cluster membership probability for individual stars is not available. 
Keeping this in  mind,  we  introduce another polarimetric cluster membership
probability (${Mp_{pol}}^a$), which can  be easily estimated without prior  proper-motion 
cluster membership information. The new cluster membership probability ${Mp_{pol}}^a$  is
estimated  in  a similar  way  as for ${Mp_{pol}}^p$ , but 
the mean  values and ranges of the Stokes vectors $q$ and $u$ are calculated considering all the stars available 
in a particular cluster.

\begin{table}
\begin{minipage}{300mm}
\caption{Reddening and scatter of the clusters}
\label{list_s}
\begin{tabular}{lllll}
\hline
{\hspace{-1mm}Cluster} & {\hspace{-1mm}$E(B-V)$} & {\hspace{-4mm} $\Delta E(B-V)$} & $Scatt_b$ &{\hspace{1mm} $Scatt_a$} \\
{\hspace{2mm}(1)}&{\hspace{4mm}(2)} & {\hspace{3mm}(3)} &{\hspace{1mm} (4)} &{\hspace{5mm}(5)} \\
\hline
{\hspace{-2mm}Hogg 15}  & {\hspace{-2mm}1.16 $\pm$ 0.03} &{\hspace{1mm} 0.20} &{\hspace{-4mm}1.02 $\pm$ 0.34}&{\hspace{-1mm} 0.64 $\pm$ 0.28} \\
{\hspace{-2mm}NGC 6611} & {\hspace{-2mm}0.85 $\pm$ 0.05} &{\hspace{1mm} 0.63} &{\hspace{-4mm}1.74 $\pm$ 0.44}&{\hspace{-1mm} 1.39 $\pm$ 0.43} \\
{\hspace{-2mm}NGC 5606} & {\hspace{-2mm}0.50 $\pm$ 0.05} &{\hspace{1mm} 0.32} &{\hspace{-4mm}1.02 $\pm$ 0.28}&{\hspace{-1mm} 1.16 $\pm$ 0.47} \\
{\hspace{-2mm}NGC 6231} & {\hspace{-2mm}0.46 $\pm$ 0.05} &{\hspace{1mm} 0.28} &{\hspace{-4mm}0.72 $\pm$ 0.14}&{\hspace{-1mm} 0.73 $\pm$ 0.14} \\
{\hspace{-2mm}NGC 5749} & {\hspace{-2mm}0.42 $\pm$ 0.04} &{\hspace{1mm} 0.13} &{\hspace{-4mm}0.34 $\pm$ 0.19}&{\hspace{-1mm} 0.34 $\pm$ 0.19} \\
{\hspace{-2mm}NGC 6250} & {\hspace{-2mm}0.33 $\pm$ 0.05} &{\hspace{1mm} 0.28} &{\hspace{-4mm}0.85 $\pm$ 0.49}&{\hspace{0mm} \ \ \ \ \ --- }\\    
\hline
\end{tabular}
\end{minipage}
\end{table}

\section{Scatter of Stokes vectors}
In cluster membership studies we consider a star to be a member of a particular cluster if
the cluster membership probability is $> 50\%$, and consider it a non-member star otherwise.
To present  the difference  in polarization between  proper-motion
member  and   
non-member stars,  we  plot the  Stokes
vectors $u(\%)$ and $q(\%)$ for all six  open clusters
in Figure $\ref{fig_uq}$,  in decreasing  order of  reddening. The
filled circles represent the proper-motion member stars and the open circles
represent the proper-motion non-member stars.

In Section 2, it was stated that the member stars of a particular cluster
should show similar interstellar polarization and  position angle if their light outputs   
encounter the same amount of dust grains and homogeneous magnetic  field, and they are all
located  at nearly the same distances. If all the member stars satisfy these conditions  
then we could expect to observe a clustering of all the member stars  in a $u(\%)$ versus $q(\%)$ plot.

We could easily infer from Figure $\ref{fig_uq}$ that the clustering of the member stars 
is very low in NGC 6611 while it is high in NGC 5749, compared with other clusters, or that the scatter of the member 
stars is very high in NGC 6611 and low in NGC 5749. 
However, by visual inspection it is very difficult to quantify the 
scatter of the member stars in the different clusters. 
Therefore, we have estimated the scatter $(Scatt_b)$ and respective errors
of the member stars in the different clusters, listed in 
the fourth column of Table $\ref{list_s}$. $Scatt_b$ is estimated from the square root of the sum of 
the squares of the Stokes vectors' standard deviations.
It is clear from $Scatt_b$  and  Fig. 1  that  the scatter 
of the  member stars in Hogg 15, NGC 6611 and NGC 5606 is high compared to that in  NGC 6231,  
NGC 5749 and NGC 6250.
In the case of clusters NGC 6611, NGC 5606, NGC 6231 and NGC 5749, it is also found that
$Scatt_b$  follows  the cluster's average  reddening,
i.e., the scatter is high towards  higher reddening and vice  versa. But, the clusters
Hogg 15 and  NGC 6250 do not follow the trend of reddening and
scatter  followed  by the  other  four  clusters. 
Further investigation is necessary for make a more precise conclusion.

\begin{figure}
\begin{center}
\includegraphics[scale = .45, trim = 10 10 10 30, clip]{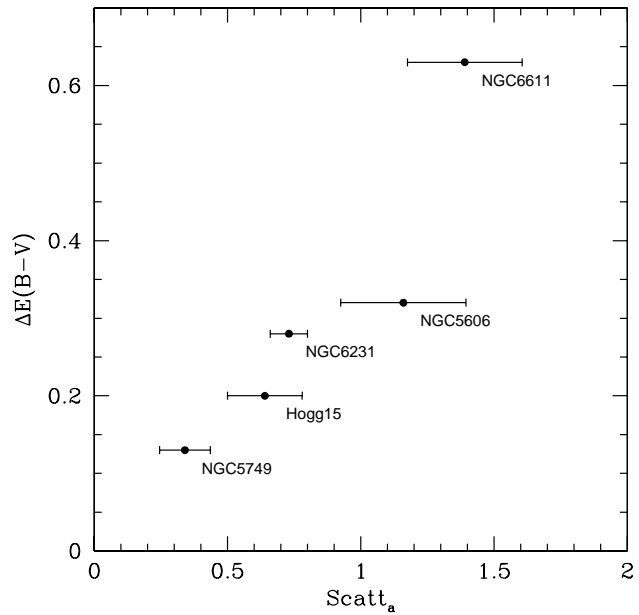}%
\hspace*{10pt}
\caption[scatt. plots of different clusters]{ \small Scatter 
($Scatt_a$) versus differential intra-cluster reddening ($\Delta E(B-V)$) 
plot for different clusters}
\label{fig_scatt} 
\end{center}
\end{figure}

\section{Stars having sources of intrinsic polarization}

It has already discussed in Section 2 that our method 
is  based on   interstellar  polarization  data,
which are extrinsic  in nature and  not variable in  time.
However, there is a chance that some stars in our sample may be sources of
intrinsic polarization. 
Hence, we have  to exclude all those stars from our sample
which are sources  of  intrinsic polarization  or  produce
variable polarization, $e.g.$, young stellar
objects, variable stars etc.

In  the cluster Hogg 15, the proper-motion  cluster membership  probability  and  
polarization data  are available for $17$  stars, and $9$  of them have
proper-motion cluster membership  probabilities greater  than  $50\%$.  According 
to the polarimetric study by
Orsatti $\etal$1998, the proper-motion member stars $\#14$ and $\#3$ (HD 311884) are both probable  
sources of
intrinsic polarization (Feinstein $\etal$1971). 
The  Wolf--Rayet star HD 311884 is a  short-period  binary with
strongly variable polarization,  modulated by the stellar orbits (Moffat
$\etal$1990).  The  proper-motion member stars  $\#$16 and  $\#$4  are both  variable in
nature,   one  being a T-Tauri star  and  other a Wolf--Rayet star (Moffat 1974).

The proper-motion cluster membership probability and  polarization data
are available  for 39 stars  in the cluster NGC 6611  and only 15 of them  have
proper-motion cluster membership  probabilities  greater than  $50\%$.  
According to  Martayan $\etal$2008, the proper-motion member stars $\#$175, $\#$313 and $\#$25  are binary in nature,
and $\#$503 is  a pre-main sequence star.
The  polarimetric study of the cluster NGC
6611  reveals that  the member star  $\#$166 is
a source of intrinsic polarization (Orsatti $\etal$2000).

In the cluster NGC 5606, $19$  stars  have  an available  proper-motion
cluster membership probability and polarimetric data, and $13$ of them have
proper-motion cluster membership probabilities  greater  than  $50\%$.  
According  to the polarimetric study  of the cluster NGC 5606, the proper-motion member 
stars $\#$1, $\#$2,
$\#$12, $\#$13, $\#$14, $\#$17 and $\#$36  are probable sources of intrinsic
polarization (Orsatti $\etal$2007).

\begin{table}
\centering
\begin{minipage}{250mm}
\caption{Membership probability for the stars in the sample}\label{list_mem}
\begin{tabular}{lllll}
\hline
ID & {\hspace {-3mm} $E(B -V)$(mag)} & {\hspace {-3mm}$Mp_{prop}$$(\%)$} & 
{\hspace {-6mm}${Mp_{pol}}^p$$(\%)$} & {\hspace {-3mm}${Mp_{pol}}^a$$(\%)$}  \\
(1)& {\hspace {3mm}(2)} & {\hspace {0mm}(3)} & {\hspace {-1mm} (4)} & {\hspace {0mm}(5)} \\
\hline
&&\ \ \ Hogg 15 && \\
\hline
05 & \ \  1.24 &\   81 &  66  &\  75 \\
08 & \ \  1.42 &\   81 &  68  &\  69 \\
10 & \ \  1.19 &\   86 &  70  &\  80 \\
17 & \ \  0.67 &\   82 &  71  &\  71 \\
23 & \ \  1.05 &\   68 &  60  &\  74 \\
02 & \ \  1.12 &\   47 &  74  &\  83 \\
06 & \ \  1.13 &\    0 &  72  &\  76 \\
11 & \ \  0.00 &\    0 &  51  &\  58 \\
15 & \ \  1.33 &\    0 &  81  &\  81 \\
19 & \ \  1.09 &\   14 &  94  &\  91 \\
20 & \ \  1.09 &\    0 &  58  &\  66 \\
21 & \ \  1.25 &\   24 &  55  &\  62 \\
22 & \ \  0.99 &\   29 &  35  &\  54 \\
\hline
&&\ \ \ NGC 6611 && \\
\hline
150 & \ \  0.76 & \    88&  90 & \  94\\
166 & \ \  0.88 & \    98&  76 & \  82\\
175 & \ \  1.16 & \    97&  75 & \  76\\
223 & \ \  0.85 & \    80&  77 & \  79\\
313 & \ \  0.72 & \    89&  90 & \  94\\
351 & \ \  0.71 & \    91&  75 & \  76\\
411 & \ \  0.10 & \    88&  79 & \  84\\
503 & \ \  0.80 & \    81&  51 & \  61\\
025 & \ \ \ \---- &  \ 53&  53 & \  55\\
231 & \ \  1.01 & \    62&  55 & \  59\\
307 & \ \ \ \---- &  \ 60&  67 & \  74\\
343 & \ \  1.11 & \    72&  86 & \  86\\
367 & \ \  0.54 & \    69&  85 & \  89\\
371 & \ \ \ \---- &  \ 58&  63 & \  70\\
444 & \ \  1.06 & \    79&  84 & \  84\\
197 & \ \  0.77 & \    3&   70 & \  77\\
205 & \ \  0.79 & \    40&  79 & \  85\\
254 & \ \  0.73 & \    44&  93 & \  96\\
259 & \ \  1.00 & \    8&   76 & \  76\\
275 & \ \  0.72 & \    12&  86 & \  86\\
280 & \ \  0.73 & \    23&  83 & \  88\\
296 & \ \ \ \---- &  \ 3&   61 & \  68\\
297 & \ \  0.92 & \    23&  83 & \  83\\
301 & \ \  0.95 & \    7&   97 & \  98\\
311 & \ \  0.76 & \    26&  84 & \  89\\
349 & \ \  0.52 & \    30&  74 & \  80\\
374 & \ \  0.56 & \    4&   80 & \  85\\
388 & \ \ \ \---- &  \ 1&   67 & \  74\\
401 & \ \  0.71 & \    25&  60 & \  64\\
402 & \ \ \ \---- &  \ 28&  48 & \  57\\
\hline
&&\ \ \ NGC 5606 && \\
\hline
01 & \ \   0.58 & \  80&  66& \  72\\ 
02 & \ \   0.56 & \  79&  69& \  89\\ 
06 & \ \   0.53 & \  64&  75& \  89\\ 
09 & \ \   0.50 & \  84&  63& \  43\\ 
12 & \ \   0.52 & \  60&  83& \  87\\ 
13 & \ \   0.35 & \  73&  71& \  85\\ 
14 & \ \   0.49 & \  51&  58& \  58\\ 
15 & \ \   0.57 & \  79&  80& \  74\\ 
17 & \ \   0.52 & \  76&  70& \  70\\ 
21 & \ \ \ \---- & \ 83&  81& \  95\\ 
36 & \ \   0.54 & \  63&  78& \  92\\ 
57 & \ \ \ \---- & \ 84&  86& \  86\\ 
60 & \ \   0.48 & \  55&  73& \  67\\ 
07 & \ \   0.51 & \  37&  70& \  84\\ 
\hline                                 
\end{tabular}                    
\end{minipage}
\end{table}

\begin{table}
\centering
\begin{minipage}{450mm}
\caption{Continuation of table \ref{list_mem}}\label{list_mem1}
\begin{tabular}{lllll}
\hline
ID & {\hspace {-10mm} $E(B-V)$(mag)} & {\hspace {-3mm}$Mp_{prop}$$(\%)$} & 
{\hspace {-5mm}${Mp_{pol}}^p$$(\%)$} & {\hspace {-3mm}${Mp_{pol}}^a$$(\%)$} \\
(1)& {\hspace {0mm}(2)} & {\hspace {0mm}(3)} & {\hspace {-1mm} (4)} & 
{\hspace {0mm}(5)} \\
\hline
&&\ \    NGC 5606 && \\
\hline
10 &     0.49 &  16& 50&  70\\ 
20 &  \ \---- &  01& 77&  94\\ 
24 &  \ \---- &  04& 46&  67\\  
58 &     0.54 &  15& 68&  89\\  
59 &     0.43 &  30& 75&  93\\ 
\hline 
 &&\ \    NGC 6231 && \\
\hline
     001    &   0.45  &  90 &  82 & 84 \\
     006    &   0.45  &  85 &  88 & 89 \\
     034    &   0.46  &  81 &  68 & 74 \\
     105    &   0.56  &  84 &  84 & 89 \\
     110    &   0.52  &  86 &  94 & 91 \\
     112    &   0.50  &  88 &  94 & 95 \\
     161    &   0.51  &  91 &  75 & 77 \\
     166    &   0.51  &  87 &  69 & 71 \\
     184    &   0.46  &  82 &  97 & 96 \\
     189    &   0.45  &  83 &  87 & 89 \\
     194    &   0.46  &  90 &  95 & 96 \\
     224    &   0.52  &  88 &  92 & 90 \\
     232    &   0.52  &  88 &  89 & 93 \\
     238    &   0.47  &  84 &  93 & 91 \\
     248    &   0.47  &  88 &  66 & 71 \\
     259    &   0.45  &  92 &  86 & 89 \\
     261    &   0.46  &  90 &  92 & 93 \\
     266    &   0.44  &  89 &  77 & 78 \\
     272    &   0.41  &  87 &  88 & 89 \\
     286    &   0.43  &  91 &  78 & 80 \\
     289    &   0.43  &  89 &  71 & 73 \\
CPD417733   & \ \---- &  91 &  94 & 90 \\
     016    & \ \---- &  52 &  89 & 90 \\
     080    &   0.60  &  63 &  72 & 75 \\
     287    &   0.45  &  58 &  66 & 69 \\
     070    &   0.54  &  0  &  61 & 67 \\
     073    &   0.64  &  0  &  81 & 86 \\
     102    &   0.45  &  0  &  90 & 94 \\
     220    &   0.47  &  0  &  85 & 86 \\
     253    &   0.46  &  23 &  86 & 91 \\
     254    &   0.47  &  0  &  95 & 97 \\
     290    &   0.53  &  0  &  85 & 86 \\
 HD152233   & \ \---- &  0  &  89 & 90 \\
 HD152235   & \ \---- &  0  &  85 & 89 \\
 HD152248   & \ \---- &  0  &  80 & 82 \\
\hline 
 &&\ \   NGC 5749 && \\
\hline
 15 & \----&  53 & 84 & 83 \\
 46 & \----&  57 & 94 & 87 \\
 72 & \----&  53 & 88 & 82 \\
 22 & \----&  44 & 70 & 89 \\
 23 & \----&  31 & 77 & 78 \\
 25 & \----&  14 & 57 & 62 \\
 26 & \----&  49 & 53 & 66 \\
 28 & \----&  43 & 99 & 87 \\
 39 & \----&  14 & 36 & 53 \\
 40 & \----&  21 & 72 & 81 \\
 31 & \----&   2 & 48 & 55 \\
 82 & \----&   3 & 37 & 53 \\
 75 & \----&  27 & 92 & 81 \\
 77 & \----&  49 & 76 & 77 \\
 38 & \----&  46 & 90 & 88 \\
\hline                               
\end{tabular}                    
\end{minipage}
\end{table}

The  proper-motion cluster  membership  probability  and   polarimetric  data are available  for  $35$  stars  in  the cluster NGC 6231, and  $25$  of them have
proper-motion membership  probabilities greater than  $50\%$. The proper-motion member
star $\#$CPD-417733 is a short-period binary (Sana $\etal$ 2008). 
The polarimetric study of the cluster NGC 6231 reveals that the  proper-motion non-member stars
$\#$70, $\#$73, $\#$220 and $\#$254 are probable  sources of intrinsic
polarization (Feinstein $\etal$2003).

Only $15$ stars have an available proper-motion  cluster membership probability
and  polarimetric data  in the cluster NGC  5749  (Vergne $\etal$2007, Dias  $\etal$2006). 
Of the $15$ stars,  only $3$  have proper-motion membership probabilities
greater than  $50\%$. The proper-motion non-member star $\#$75 is a probable
source of intrinsic polarization (Vergne $\etal$2007).
     
In the cluster NGC 6250 proper-motion cluster membership probability and polarization
data are available for $33$  stars, and only  $3$ of  them have  proper-motion membership
probabilities greater  than $50\%$. According to the polarimetric study 
of the cluster NGC 6250, the proper-motion member  stars $\#$11 and $\#$35, and non-member
stars $\#$13, $\#$18, $\#$19, $\#$37 are 
probable sources of intrinsic polarization (Feinstein $\etal$2008).  

\begin{table*}
\centering
\begin{minipage}{150mm}
\caption{Results for NGC 6231 }\label{list_mem2}
\begin{tabular}{lllllllll}
\hline
ID & {\hspace {-9mm} $E(B-V)$(mag)} & {\hspace {-3mm}$Mp_{prop}$$(\%)$} & {\hspace {-3mm}${Mp_{pol}}^p$$(\%)$} 
&{\hspace {-3mm} ${Mp_{pol}}^a$$(\%)$} & $P_{max}$$\pm$ $\epsilon$ $(\%)$ & {\hspace {+3mm}$\sigma_{1}$} 
& $\lambda_{max}$$\pm$ $\epsilon$ ($\mu m$)  & {\hspace {2mm}$\overline{\epsilon}$}\\
(1) & (2) & (3) & (4) & (5) & {\hspace {+3mm}(6)} & {\hspace {+3mm}(7)} & {\hspace {+3mm}(8)} & {\hspace {+1mm}(9)} \\
\hline                                      
     001    &   0.45  &  90 &  82 & 84 & 0.846 $\pm$ 0.112 & 1.062 & 0.553 $\pm$ 0.189 &  18.9\\
     006    &   0.45  &  85 &  88 & 89 & 0.355 $\pm$ 0.006 & 0.169 & 0.615 $\pm$ 0.027 &  10.4\\
     034    &   0.46  &  81 &  68 & 74 & 1.511 $\pm$ 0.077 & 0.743 & 0.656 $\pm$ 0.064 &  12.2\\
     105    &   0.56  &  84 &  84 & 89 & 1.175 $\pm$ 0.027 & 0.538 & 0.493 $\pm$ 0.023 &   2.3\\
     110    &   0.52  &  86 &  94 & 91 & 0.476 $\pm$ 0.075 & 0.932 & 0.427 $\pm$ 0.117 &   6.1\\
     112    &   0.50  &  88 &  94 & 95 & 0.549 $\pm$ 0.018 & 0.463 & 0.536 $\pm$ 0.031 &   4.1\\
     161    &   0.51  &  91 &  75 & 77 & 0.441 $\pm$ 0.141 & 1.612 & 0.321 $\pm$ 0.098 &   1.9\\
     166    &   0.51  &  87 &  69 & 71 & 0.699 $\pm$ 0.031 & 0.209 & 0.439 $\pm$ 0.030 &   2.1\\
     189    &   0.45  &  83 &  87 & 89 & 0.558 $\pm$ 0.131 & 0.629 & 0.377 $\pm$ 0.101 &   3.0\\
     194    &   0.46  &  90 &  95 & 96 & 1.018 $\pm$ 0.371 & 0.987 & 0.205 $\pm$ 0.041 &   2.0\\
     224    &   0.52  &  88 &  92 & 90 & 0.345 $\pm$ 0.027 & 0.624 & 0.507 $\pm$ 0.079 &   6.0\\
     232    &   0.52  &  88 &  89 & 93 & 0.911 $\pm$ 0.018 & 0.220 & 0.484 $\pm$ 0.013 &   2.0\\
     238    &   0.47  &  84 &  93 & 91 & 0.407 $\pm$ 0.028 & 0.451 & 0.549 $\pm$ 0.083 &   7.9\\
     248    &   0.47  &  88 &  66 & 71 & 1.614 $\pm$ 0.043 & 0.350 & 0.468 $\pm$ 0.029 &   9.1\\
     259    &   0.45  &  92 &  86 & 89 & 0.891 $\pm$ 0.035 & 0.566 & 0.472 $\pm$ 0.031 &   1.3\\
     261    &   0.46  &  90 &  92 & 93 & 0.588 $\pm$ 0.017 & 0.257 & 0.450 $\pm$ 0.021 &   1.4\\
     266    &   0.44  &  89 &  77 & 78 & 0.396 $\pm$ 0.024 & 0.709 & 0.589 $\pm$ 0.069 &   5.6\\
     272    &   0.41  &  87 &  88 & 89 & 0.592 $\pm$ 0.063 & 0.643 & 0.382 $\pm$ 0.057 &   5.1\\
     286    &   0.43  &  91 &  78 & 80 & 0.533 $\pm$ 0.118 & 0.864 & 0.348 $\pm$ 0.098 &   8.6\\
     289    &   0.43  &  89 &  71 & 73 & 0.731 $\pm$ 0.032 & 0.788 & 0.439 $\pm$ 0.034 &   0.5\\
 CPD417733  & \ \---- &  91 &  94 & 90 & 0.444 $\pm$ 0.034 & 0.399 & 0.852 $\pm$ 0.095 &   5.6\\
     016    & \ \---- &  52 &  89 & 90 & 0.371 $\pm$ 0.023 & 0.338 & 0.542 $\pm$ 0.065 &   2.3\\
     080    &   0.60  &  63 &  72 & 75 & 1.036 $\pm$ 0.031 & 0.577 & 0.551 $\pm$ 0.036 &   8.6\\
     287    &   0.45  &  58 &  66 & 69 & 0.830 $\pm$ 0.012 & 0.207 & 0.474 $\pm$ 0.011 &   0.6\\
     070    &   0.54  &  0  &  61 & 67 & 1.616 $\pm$ 0.274 & 5.933 & 0.627 $\pm$ 0.192 & 367.0\\
     073    &   0.64  &  0  &  81 & 86 & 1.953 $\pm$ 0.380 & 7.997 & 0.725 $\pm$ 0.220 & 199.5\\
     102    &   0.45  &  0  &  90 & 94 & 0.847 $\pm$ 0.011 & 0.236 & 0.581 $\pm$ 0.015 &  16.2\\
     220    &   0.47  &  0  &  85 & 86 & 0.657 $\pm$ 0.024 & 0.360 & 0.506 $\pm$ 0.035 &  14.4\\
     253    &   0.46  &  23 &  86 & 91 & 0.985 $\pm$ 0.044 & 0.524 & 0.480 $\pm$ 0.041 &   7.0\\
     254    &   0.47  &  0  &  95 & 97 & 0.928 $\pm$ 0.078 & 0.867 & 0.460 $\pm$ 0.091 &  33.5\\
     290    &   0.53  &  0  &  85 & 86 & 0.666 $\pm$ 0.012 & 0.163 & 0.469 $\pm$ 0.014 &   6.0\\
 HD152233   & \ \---- &  0  &  89 & 90 & 0.609 $\pm$ 0.066 & 0.926 & 0.514 $\pm$ 0.161 &   7.4\\
 HD152235   & \ \---- &  0  &  85 & 89 & 1.020 $\pm$ 0.021 & 0.386 & 0.465 $\pm$ 0.015 &   1.5\\
 HD152248   & \ \---- &  0  &  80 & 82 & 0.562 $\pm$ 0.029 & 0.462 & 0.533 $\pm$ 0.056 &   6.0\\
\hline 
\end{tabular}                    
\end{minipage}
\end{table*}

\begin{figure*}
\begin{center}
\includegraphics[scale = .49, trim = 10 40 80 30, clip]{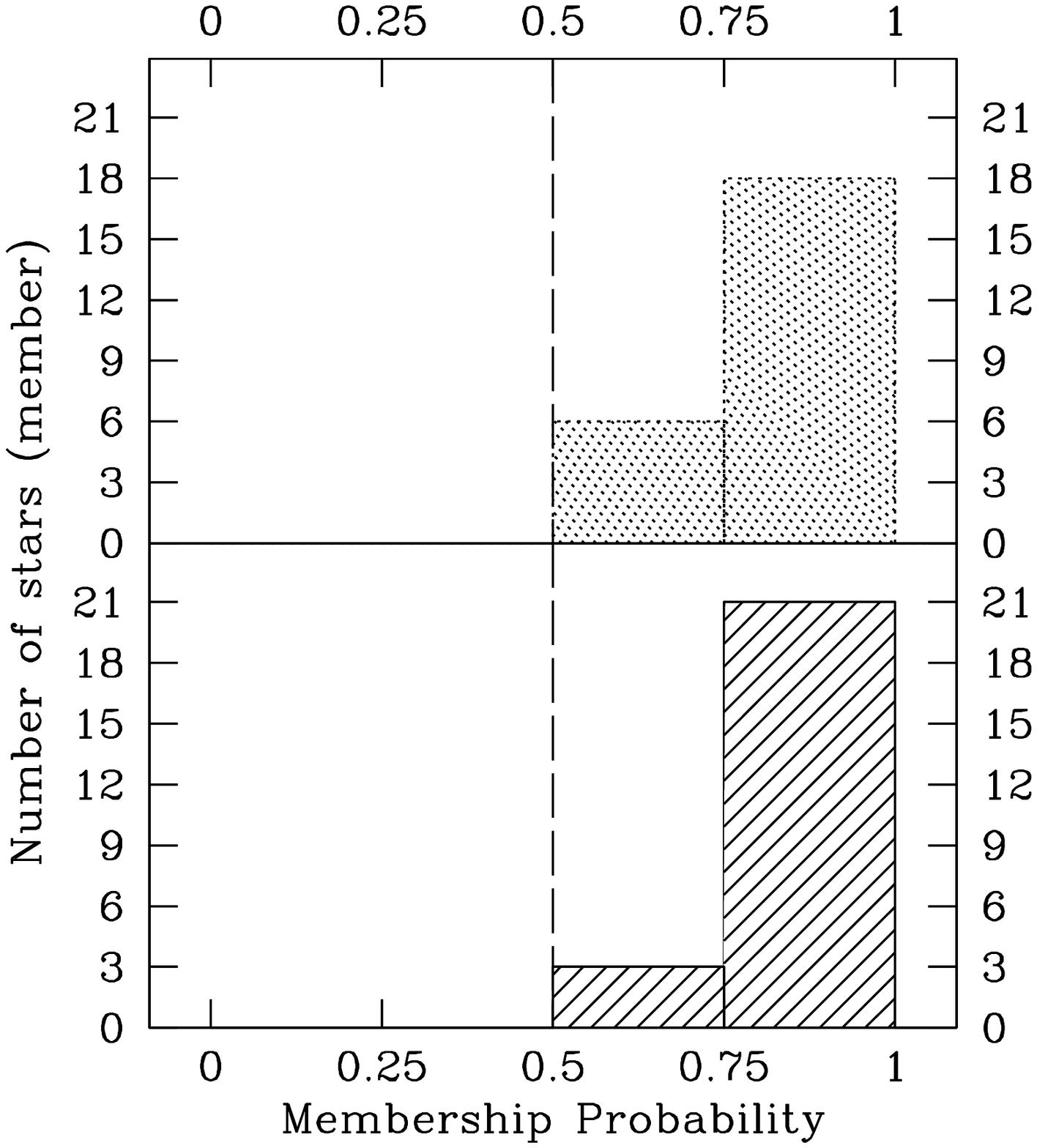}%
\hspace*{10pt}
\includegraphics[scale = .49, trim = 10 40 80 30,clip]{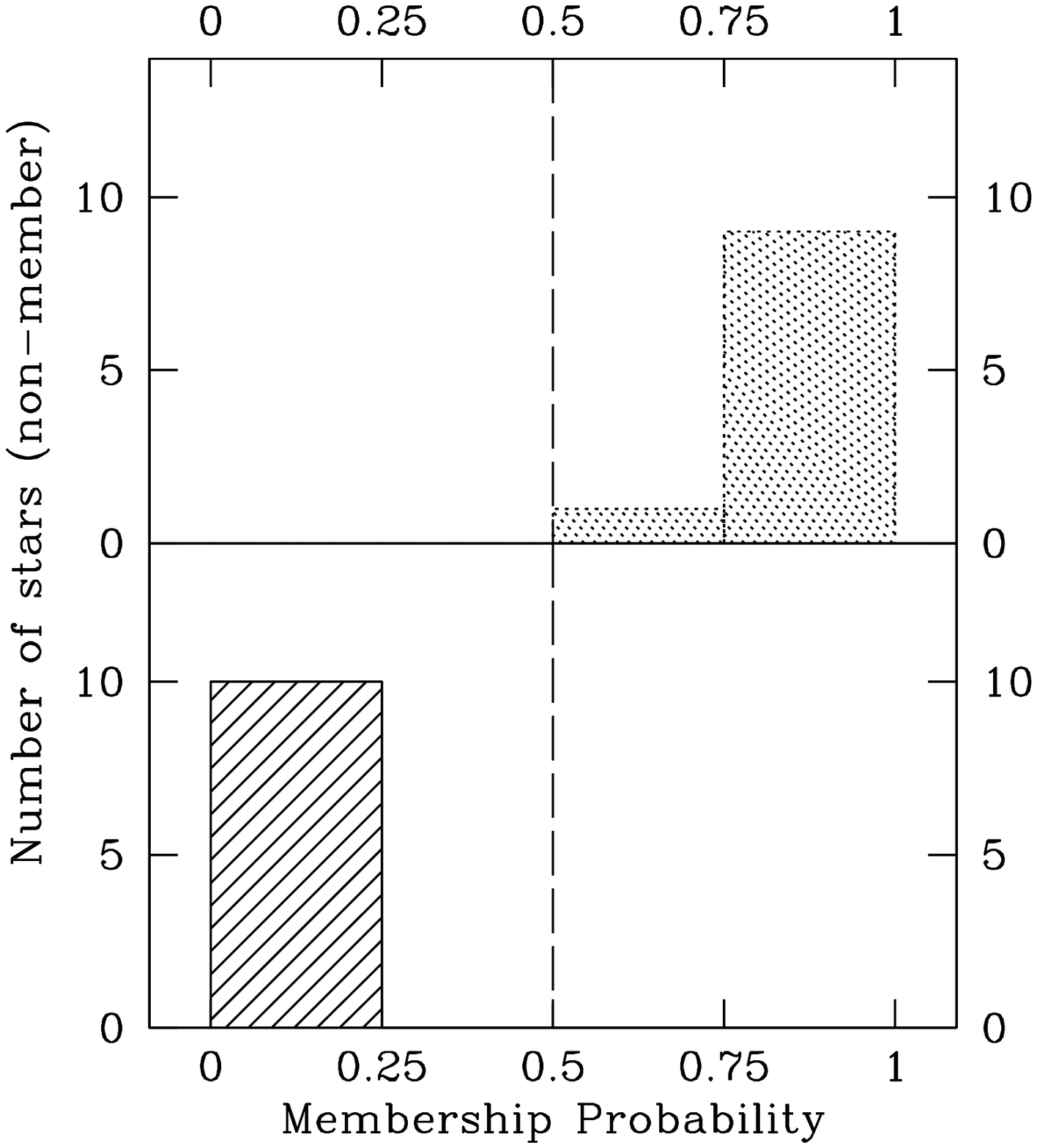}\\
\hspace*{10pt}
\caption[comparison of membership probability]{\small Comparison of cluster 
membership probability for the stars belonging to the cluster NGC 6231,
determined by two different techniques based on proper-motion and interstellar polarization, respectively.  
The left hand side histogram  shows a comparison between the proper-motion and polarization 
cluster membership probability for the proper-motion member stars and the right hand side 
histogram shows a comparison for the proper-motion non-member stars.
 The dotted and lined  histograms represent the 
results determined by the polarization and proper-motion cluster membership techniques, respectively.}

\label{comp_mem} 
\end{center}
\end{figure*}

The above mentioned  proper-motion member  stars in different clusters are
shown  in Fig.$\ref{fig_uq}$, and are 
probable  sources  of  intrinsic and/or variable polarization.
After excluding all these stars from the
sample of proper-motion member stars, the scatter is
similar to the case when they are included  in the clusters NGC 5606, NGC 6231 and NGC 5749. However, the scatter
decreases in the clusters Hogg 15, NGC  6611 and NGC 6250.
The estimated scatter $(Scatt_a )$ and the respective errors of  the different clusters  
after  exclusion of stars that are probable sources  of  intrinsic polarization  or  produce
variable polarization are given in the fifth column
of Table $\ref{list_s}$. 

\section{Polarimetric cluster membership probability for stars}

There are very few open clusters with
available interstellar polarization data and a proper-motion cluster membership
probability. After an extensive database survey we found 
six open clusters for analysis. Of these six clusters, two of them, NGC 5749 
and NGC 6250, have only three stars with
an available proper-motion cluster membership probability.
After excluding the stars having probable 
sources of intrinsic/variable polarization only one proper-motion member star is left
in the cluster NGC 6250. So, it is  impossible
to comment about the scatter $Scatt_a$ in the cluster NGC 6250. 

If we assume that the scatter $Scatt_a$ of the Stokes vectors $u(\%)$ and $q(\%)$ 
depends on the differential intra-cluster reddening $\Delta E(B-V)$, then 
$Scatt_a$  should increase with increasing $\Delta E(B-V)$.
The values of $\Delta E(B-V)$ for all six open clusters are given in
the third  column of  Table $\ref{list_s}$. Of these  six
clusters NGC 5749 has the lowest value of $\Delta E(B-V)$ $\sim$ 0.13
mag and NGC  6611 has the  highest value of $\Delta
E(B-V)$ $\sim$ 0.63 mag.
A  $Scatt_a$  versus $\Delta  E(B-V)$ plot is shown in Figure $\ref{fig_scatt}$. 
The cluster NGC 6250  is not included in
the plot because only one member star is left in the sample. It is clear from  the plot  
that $Scatt_a$ increases almost proportionally with $\Delta E(B-V)$ and all the 
five clusters follow a similar trend.  
Therefore, the  above analysis suggests that the scatter of the Stokes vectors $u(\%)$ and $q(\%)$
of the proper-motion member stars depends  on 
the intra-cluster differential reddening  $\Delta E(B-V)$
of different clusters.

The  cluster membership  probabilities $Mp_{prop}$, ${Mp_{pol}}^p$ and ${Mp_{pol}}^a$
of different stars belonging to a particular 
cluster  are given in  the third, fourth and fifth  columns  of  Table
$\ref{list_mem}$ and  Table $\ref{list_mem1}$. $Mp_{prop}$ is the  
cluster membership  probability   based  on   the
proper-motion study given in the third column.
${Mp_{pol}}^p$ and  ${Mp_{pol}}^a$ are
the cluster membership  probabilities based on the interstellar polarimetric data given in 
the fourth and fifth columns. Details about the estimation of the cluster membership probabilities 
${Mp_{pol}}^p$ and ${Mp_{pol}}^a$
were discussed in Section 3. The estimations of
${Mp_{pol}}^p$ and ${Mp_{pol}}^a$ are based on 
interstellar polarimetric data, though in the case of the estimation of ${Mp_{pol}}^p$, the proper-motion 
cluster membership probability is used as a reference. From Table $\ref{list_mem}$ and $\ref{list_mem1}$  
it is found that both 
${Mp_{pol}}^p$ and
${Mp_{pol}}^a$  are comparable, and the cross-correlation coefficient(r) between them is 0.92.

Using a cluster membership criterion (membership probability $> 50\%$ for member stars and $< 50\%$
for non-member stars)
we divided the stars  into two groups, members and non-members. It is found that for 
proper-motion member stars both the polarimetric
cluster membership probabilities ${Mp_{pol}}^p$  and  ${Mp_{pol}}^a$ follow
the proper-motion cluster membership probability $Mp_{prop}$, and
fall into the same group, $i.e.$, member stars group. The cross-correlation coefficients (r) between $Mp_{prop}$ and ${Mp_{pol}}^p$, and 
$Mp_{prop}$ and ${Mp_{pol}}^a$ are 0.70 and 0.72, respectively.
But for proper
motion non-member stars the polarimetric membership probabilities ${Mp_{pol}}^p$ and ${Mp_{pol}}^a$ do not follow
the proper-motion cluster membership probability $Mp_{prop}$. 
The polarimetric cluster membership probabilities ${Mp_{pol}}^p$  and  ${Mp_{pol}}^a$ should fall in the 
non-member stars group,  but they fall into the opposite group.

Figure $\ref{comp_mem}$ shows a comparison between proper-motion
and polarization cluster membership probabilities for the stars belonging to the cluster NGC 6231. 
The left hand side histograms show a comparison between ${Mp_{prop}}$ and ${Mp_{pol}}^a$ 
for the proper-motion member stars, and the right hand side 
histograms show a comparison for the proper-motion non-member stars.
 The dotted and lined  histograms present the 
cluster membership probabilities determined by the polarization and proper-motion techniques, respectively.
It can be easily inferred from Figure $\ref{comp_mem}$ that for the proper-motion member 
stars, the polarimetric cluster membership probabilities
are  in good agreement with the
proper-motion cluster membership probability
whereas for the proper-motion non-member
stars the polarimetric cluster membership probabilities totally disagree with
the proper-motion cluster membership probability.

\section{Open cluster NGC 6231}
Let us consider the cluster NGC 6231 to 
study the consistency and validity of the polarimetric membership probability  in detail.
The  reddening $E(B-V)$ and  membership probabilities  $Mp_{prop}$, ${Mp_{pol}}^p$ and
${Mp_{pol}}^a$ of the different individual stars are  given  in the  second,  third,  fourth and fifth  columns
of Table $\ref{list_mem2}$.
The  maximum  value  of  polarization ($P_{max}$)  and
wavelength ($\lambda_{max}$)  are given  in  the sixth and eighth  columns of
Table $\ref{list_mem2}$. 
$\lambda_{max}$ and $P_{max}$ are both
 functions  of  the optical properties  and characteristics  of the particle
size distribution  of the aligned dust grains  (McMillan 1978, Wilking
$\etal$1980).  The value of $\lambda_{max}$  and $P_{max}$  are calculated
by  fitting  the  observed interstellar polarization data in $U$,  $B$,  $V$, $R$  and  $I$
band-pass filters using the standard  Serkowski's polarization  law (Feinstein et al. 2003, Serkowski
1973):
\begin{equation}
P_{\lambda}/    P_{max}   =\    exp    \left[-\   k    \   ln^{2}    \
(\lambda_{max}/\lambda) \right]
\end{equation}
and  adopting  the   parameter  $K  =  1.66$ $\lambda_{max}$ $\pm$  $0.01$
(Whittet et al. 1992).

The stars $\#034$, $\#194$, $\#261$ and
$\#253$ have a value of $E(B-V)$ $\simeq$ 0.46 mag, which is nearly equal
to the average value of cluster reddening.
The proper-motion cluster membership probability
 $Mp_{prop}$ and polarimetric
cluster membership probability ${Mp_{pol}}^a$ for these four stars are $81\%$,
$90\%$, $90\%$ and $23\%$, and  $74\%$, $96\%$, $93\%$ an d $91\%$, respectively.
$Mp_{prop}$  for
star $\#253$ is $23\%$  and ${Mp_{pol}}^a$  for the
same star is $91\%$. 
So, according to proper-motion technique,
star $\#253$ is a non-member, but according to the polarimetric technique it is a member star
of the cluster NGC 6231.  
The same trend is observed for the  remaining proper-motion non-member
stars. 
For example, the  stars $\#70$, $\#73$, $\#102$, $\#220$, $\#254$ and
$\#290$ have reddening $E(B-V)$ values of 0.54, 0.64, 0.45, 0.47, 0.47 and 0.53
mag. The proper-motion cluster membership probability $Mp_{prop}$ for all these stars is $00.0\%$, 
but the  polarimetric
cluster membership probability ${Mp_{pol}}^a$ is
$67\%$, $86\%$, $94\%$, $86\%$, $97\%$ and $86\%$, respectively.
So, it is clear from
the above discussions  that in the proper-motion non-member regime the 
methods do not  agree.
The main reason behind this may be the
fundamental differences between the two techniques: $Mp_{prop}$ is based on the stellar proper
motion in space and ${Mp_{pol}}^a$ is based on the interstellar linear polarization, $i.e.$,
the selective extinction  of star light by  aligned asymmetric
dust  grains present  in the  interstellar  medium along  the line  of
sight.

Since reddening and polarization both originate from similar physical mechanisms, it is expected that the stars $\#238$ and $\#248$ should have the same value of polarimetric
cluster membership probability 
as they both have the  same value of reddening $\simeq 0.47$ mag. 
However, experimentally, this is found to not be the case.
We have already  stated in Section 2  that the
correlation between normal and selective extinction by the
asymmetric aligned dust grains cannot be maintained for all cases due to
the variations of grain size and alignment efficiency.  The entire
population  of dust grains  which  are  not  aligned  and  elongated  may
contribute only to reddening and not  to polarization. 
The  value of
$\lambda_{max}$ for  star $\#238$ is  0.55 $\mu m$ and  for star
$\#248$, it is 0.47 $\mu m$. From these values,  
it is clear 
that the  line of sight for both stars is not  populated by 
grains of similar size, generated by a similar method.
Hence, it  is
possible to  have different interstellar linear polarizations as well  as different
${Mp_{pol}}^a$ for different stars even though they all have the same value of reddening.

\section{Discussion}

From the above results and analysis it is found that our method of estimating the cluster 
membership probability using linear polarimetric data  is applicable only 
to proper-motion member stars. We can apply this technique to estimate the 
cluster membership probability for known member stars having no  membership probability.
However, other  techniques are required  to  eliminate probable
non-member stars and stars  having sources of intrinsic polarization from the membership sample. 
Our technique can then be used to  determine  the cluster membership
probability  of  any  star belonging to a particular cluster. 

The dispersion of reddening/extinction of a particular star from the mean value of that specific cluster 
may  be  used to  determine  the probable
non-member stars in a cluster. Alternatively, the  cluster photometry could  be used to
identify the probable non-member stars in a cluster.
Polarimetry is also
a  very  powerful  tool  for determining  stars that have sources of  intrinsic
polarization. 
The unit  weight error of the fit  ($\sigma_1$) and 
dispersion of position  angle ($\overline{\epsilon}$) for the stars belonging to cluster NGC 6231 were determined (given in
the seventh and ninth column of Table $\ref{list_mem2}$). 
A value of $\sigma_1$, calculated for each star
during  the fitting  of Serkowski's  law,  
of less  than 1.5 due to the weighting scheme indicates that the polarization  is well
represented by Serkowski's interstellar polarization  law (Medhi $\etal$2010, 2008, 2007).
A  higher value  could  be  indicative of  the  presence of  intrinsic
polarization.  
The dispersion of the position angle
($\overline{\epsilon}$) for each star  normalized by the mean value of
the position  angle errors is another tool for  detecting stars
having 
sources of intrinsic polarization.
 
We consider seven proper-motion non-member stars from the open cluster NGC 6231 $\#70$, $\#73$, $\#102$, $\#220$, 
$\#253$, $\#254$ and $\#290$, and assume that proper-motion membership
probabilities are not available for all of them. According to the polarimetric cluster membership probability,
all of them have a membership probability of $>$ 50$\%$, $i.e.$,
all the seven proper-motion non-member stars were identified as being  member stars of the cluster NGC 6231.  
We can use the multi-band linear polarimetric data of the same stars as a supplement to identify whether
they are members or non-members. 
According to a multi-band linear polarimetric study, the dispersion of position  
angle $\overline{\epsilon}$ for all the seven stars  
is very high ($\ge6$), which implies that they are all 
probably sources of intrinsic polarization. Therefore, we could easily eliminate
them as non-member stars from our sample of polarimetric member stars. 

The European Space Agency's space mission Global Astrometric Interferometer for Astrophysics (GAIA) 
will create an extremely precise three-dimensional map of stars throughout our Milky Way galaxy and beyond.
One of the main objectives is to determine the positions, distances and annual proper motions of nearly one billion 
stars with an expected accuracy of about 7--22 $\mu$as down to 15 mag and sub-$\mu$as accuracies
at the fainter limit of nearly 20 mag (Lindegren \etal 2007). It is expected that the GAIA will provide very accurate 
membership information of all stars belongs to open clusters. Once the GAIA data is available, it 
would provide a much larger sample for cross-checking the polarimetric
method.

\section{Summary}

The findings of  the cluster membership study using a polarimetric approach   
can be summarized as follows.

We have analyzed interstellar polarimetric data for six
open clusters,  Hogg 15, NGC 6611, NGC 5606,  NGC 6231, NGC 5749
and NGC 6250, and estimated the polarimetric cluster membership probabilities
for stars belonging to a particular cluster.
The analysis suggests that the scatter of the Stokes vectors
$q(\%)$ and $u(\%)$ of the proper-motion member stars increases with the
highly  varying  intra-cluster  reddening, $\Delta E(B-V)$.

For proper-motion member stars, the polarimetric cluster membership 
probability ${Mp_{pol}}^a$ and proper-motion cluster membership
probability $Mp_{prop}$ agree.   
However, for  proper-motion non-member  stars,  the polarimetric cluster membership  probability
is in total disagreement 
with the proper-motion cluster membership probability, showing that the polarimetric method is 
inaccurate for non-member stars.  
This may be because of fundamental
differences  between  the  two  methods,  in that one is  based  on  the stellar
proper-motion in space and  other is  based  on the interstellar polarization, 
$i.e.$, the selective  extinction of  the
stellar output  by the asymmetric  aligned dust grains present  in the
line of sight.

The polarimetric cluster membership determination technique could be used to estimate 
the cluster membership probability
of any star belonging to a particular cluster if we can identify it as a probable member/non-member of that particular 
cluster using additional polarimetric and 
photometric information for that star, such as the maximum value of the wavelength $\lambda_{max}$, the unit weight error of the
fit $\sigma_1$, the dispersion in the polarimetric position angles $\overline{\epsilon}$, the reddening $E(B-V)$
or the differential intra-cluster reddening $\Delta E(B-V)$. 
This technique could also be used to  
estimate the cluster membership probability for the
known member stars with unknown membership probability as well as to resolve 
disagreements about membership between different proper-motion surveys (Dias $\etal$2006, Baumgardt $\etal$2000, 
Belikov $\etal$1999, Tucholke $\etal$1986, Berger 1982).

\section*{Acknowledgments}

The authors wish to thank the referee for his constructive remarks which have led to
a great improvement in the clarity of the paper. It is our pleasure 
to thank Prof. Ram Sagar for his constant guidance and 
Dr. R.K.S. Yadav for useful discussions.
This research has
made use of the WEBDA data base, operated at the Institute for
Astronomy of the University of Vienna, and of IRAF, distributed by 
the National Optical Astronomy Observatories, USA.   The  author (BJM)  
would like  to  thank his wife Orchid and daughter Sanskriti for 
their support.

\section*{REFERENCES}
Ahumada A. V., Clariá J. J., Bica E., Piatti A. E., 2000, A\&AS, 141, 79 \\
Baumgardt, H., Dettbarn, C., Wielen, R. 2000, A\&AS, 146, 251 \\
Bayer C., Maitzen H. M., Paunzen E., Rode Paunzen M., Sperl M., 2000, A\&AS, 147, 99 \\
Belikov A. N., Kharchenko N. V., Piskunov A. E., Schilbach E., 1999, A\&AS, 134, 525 \\ 
Berger M.,1982, ApJ, 263, 199 \\ 
Chini R., Krugel E., 1983, A\&A 117, 289 \\ 
Claria Juan J., Lapasset E., 1992, AcA, 42, 343 \\
Cudworth Kyle M., 1997, APSC, 127, 91 \\
Cudworth Kyle M.,1986, IAUS, 109, 201 \\
Davis, L. Jr., Greenstein, Jesse L., 1951, ApJ, 114, 206 \\
de Winter D., Koulis C., The P.S., et al., 1997, A\&AS 121, 223\\
Dias W. S., Assafin M., Flório V., Alessi B. S., Líbero V., 2006, A\&A, 446, 949 \\
Feinstein Carlos, Martínez Ruben, Vergne M. Marcela, Baume Gustavo, Vazquez Ruben, 2003, ApJ, 598, 349 \\
Feinstein Carlos, Vergne M., Marcela M. Ruben, Orsatti A. Maria, 2008, MNRAS, 391, 447 \\
Feinstein Carlos, Vergne M. Marcela, Martínez Ruben, Orsatti Ana Maria, 2008, MNRAS, 391, 447 \\
Feinstein A., Marraco H. G., 1971, PASP, 83, 219 \\
Feinstein A., Ferrer S.E., 1968, PASP, 80, 410 \\
Gebel W.L., 1968, ApJ 153, 743 \\
Hillenbrand L.A., Massey P., Strom S.E., Merrill K.M., 1993, AJ 106, 1906\\
Johnson H.L., 1968, in Nebulae and Interstellar Matter, Stars and Stellar Systems VII, Middlehurst B.M. 
and Aller L.H.(eds.), p. 167 \\
Lazarian A., Goodman Alyssa A., Myers Philip C.,1997, ApJ, 490, 273 \\ 
Lindegren L., Babusiaux C., Bailer-Jones C., Bastian U., Brown A.G.A. \etal 2007, Proc. IAU Symposium, 248, 217 \\ 
Martayan C., Floquet M., Hubert A. M., Neiner C., Fremat  Y. \etal 2008, A\&A, 489, 459 \\
Medhi Biman J., Maheswar G., Pandey J. C., Tamura Motohide, Sagar R., 2010, MNRAS, 403, 1577 \\
Medhi Biman J., Maheswar G., Pandey J. C., Kumar T.S., Sagar R.,  2008, MNRAS, 388, 105 \\
Medhi Biman J., Maheswar G., Kumar B., Pandey J. C., Kumar T.S., Sagar R., 2007, MNRAS, 378, 881 \\
Moffat A. F. J., Drissen L., Robert Carmelle, Lamontagne Robert, Coziol Roger \etal, 1990, ApJ, 350, 767 \\
Moffat A. F. J., 1974, A\&A, 34, 29 \\  
Orsatti A. M., Vega E., Marraco H. G., 1998, AJ, 116, 266O \\
Orsatti A. M., Feinstein C., Vega E. I., Vergne M. M., 2007, A\&A, 471, 165O \\
Orsatti A. M., Vega E. I., Marraco H. G., 2000, A\&AS, 144, 195O \\
Piatti Andres E., Bic, Eduardo, Claria Juan J., Santos Joao F. C., Ahumada Andrea V.,2002, MNRAS, 335, 233\\
Ram Sagar, P U. Munari, K. S. de Boer, 2001, MNRAS, 327, 23 \\
R. K. S. Yadav, Ram Sagar, 2001, MNRAS, 328, 370 \\ 
Sagar R., Joshi U.C., 1979, Ap\&SS 66, 3 \\
Sana H., Rauw G., Nazé Y., Gosset E., Vreux J.M., 2006, MNRAS, 372, 661 \\
Sana H., Rauw G., Sung H., Gosset E., Vreux J.M., 2007, MNRAS, 337, 945 \\
Sana H., Gosset E., Naze Y., Rauw G., Linder N., 2008, MNRAS, 386, 447 \\
Serkowski, K., 1973, IAUS, 52, 145 \\
Sung Hwankyung, Bessell Michael S., Lee See Woo, 1998, AJ, 115, 734 \\
Turner D.G., 1994, Rev. Mex. Astron. AstrLs. 29, 163 \\
Tucholke H.J., Geffert M., The P.S.,1986, A\&AS, 66, 311 \\
Vergne M. M., Feinstein C., Martínez R., 2007, A\&A, 462, 621 \\
Vazquez Ruben A., Feinstein A., 1991, A\&AS, 87, 383 \\
Vazquez R. A., Baume G., Feinstein A., Prado P,1994, A\&AS, 106, 339 \\
Whittet D. C. B., Martin P. G., Hough J. H., Rouse M. F., Bailey J. A.\etal,  1992, ApJ, 386, 562 \\
Zacharias N., Urban S. E., Zacharias M. I., Wycoff G. L., Hall, D. M.\etal, 2004, AJ, 127, 3043 \\

\bsp

\label{lastpage}

\end{document}